\documentclass[12pt]{article}
\usepackage{epsfig}

\def\beq{\begin{equation}}
\def\eeq#1{\label{#1}\end{equation}}
\def\eeqn{\end{equation}}

\def\beqa{\begin{eqnarray}}
\def\eeqa#1{\label{#1}\end{eqnarray}}
\def\eeqan{\end{eqnarray}}

\let\bar=\overbar

\def\Dslash{\not{\hbox{\kern-4pt $D$}}}
\def\dslash{\not{\hbox{\kern-2pt $\del$}}}

\def\msb{{\bar{\ssstyle M \kern -1pt S}}}

\def\Title#1{\begin{center} {\Large {\bf #1} } \end{center}}

\def\Journal#1#2#3#4{{#1} {\bf #2}, #3 (#4)}

\def\NPB{{\em Nucl. Phys.} B}
\def\PLB{{\em Phys. Lett.}  B}
\def\PRL{\em Phys. Rev. Lett.}
\def\PRD{{\em Phys. Rev.} D}

\begin{document}

\Title{Top quark production at the Tevatron}

\bigskip\bigskip

%+\addtocontents{toc}{{\it D. Reggiano}}
%+\label{ReggianoStart}

\begin{raggedright}  

{\it Marc Besan\c con \index{Besan\c con, M.}\\
CEA-Saclay, DSM/Irfu/SPP\\
Bat.~141\\
91191 Gif sur Yvette, France}
\bigskip\bigskip
\end{raggedright}

\section{Introduction}
\label{sec:introduction}
The Tevatron is the place of the discovery of the top
quark ($t$) by both the CDF and the D0 experiments~\cite{top-discovery}.
The Tevatron worked extremely well and stopped taking data on 
September 30$^{th}$ 2011. The Tevatron delivered about 12~fb$^{-1}$
of $p \bar p$ collisions
and both the CDF and D0 experiments registered more than 10~fb$^{-1}$ of
data. The results reported in this mini-review are based on datasets
up to 8.7~fb$^{-1}$.
Doing top quark physics means covering a wide spectrum of different
subjects including studies of the $t$ (single and pair) 
production, decay and properties. The present mini-review focuses
on results which were available at the time of the 2012 FPCP 
conference on the measurements of the $t \bar t$ cross~section, 
the forward backward asymmetry, the spin correlations and the ratio of 
branching fractions as well as on single $t$ cross~section and $|V_{tb}|$
measurements. We will also report on recent searches for new resonances 
with $t \bar t$ final states. 
At the Tevatron, within the Standard Model (SM), $t \bar t$ production is
expected to occur via strong interactions namely through ${q \bar q}$ 
annihilation  (85\%) and gluon gluon ($gg$) fusion (15\%). Typical 
next-to-leading (NLO) predictions amount to
$\sigma_{NLO}(p \bar p \rightarrow t \bar t ) = 7.46^{+0.48}_{-0.67}$~pb 
for $m_{t}=172.5$~GeV~\cite{topair1} and for NNLO
$\sigma_{NNLO}(p \bar p \rightarrow t \bar t ) = 7.067^{+0.143}_{-0.232}$~(scales)~$^{+0.186}_{-0.122}$~(pdf)~pb.
for $m_{t}=173.3$~GeV~\cite{topair2}. 
On the other hand, single~$t$ production is expected to occur via 
electroweak interactions either in the s-channel (33 \%) or the t-channel 
(67 \%)
with the following SM predictions for the cross sections
$\sigma^{s-channel} = 1.05 \pm 0.07$~pb and
$\sigma^{t-channel} = 2.10 \pm 0.19$~pb
for $m_{t}=172.5$~GeV~\cite{singletop}.
The single $t$ associated production $Wt$ having a production 
cross~section of the order of 0.2~pb is too small at the Tevatron.
The $t$ decays before it hadronizes. The $t$ decays into 
a b~quark and 
an on-shell W~gauge boson ($t\rightarrow Wb$) with a branching ratio close 
to 1.
The final states corresponding to $t \bar t$ production 
are classified according to the decay of the W~gauge boson from the parent $t$.
The results reported here concentrate on the lepton+jets
channels where one W~gauge boson decays leptonically i.e. 
$W \rightarrow l \nu$ where $l= \mu$~or~$e$ (amounting to 30\% 
of all the $t \bar t$ channels) and dilepton channels where both W~decay 
leptonically (5\%). 
Other channels include all hadronic channels where both W~gauge 
bosons decay hadronically ($W \rightarrow q q'$, 45\%) 
as well as tauonic channels where both W decay
leptonically into tau leptons (20\%).
Bottom quarks are always present in the final state
and we can use b-quark identification techniques
to help for the selection of top quark events.

\section{top quark pair production}
\label{sec:topairprod}
The top quark is a unique particle. It is the heaviest of all known particle
and it decays before hadronizing. Studying $t \bar t$ production provides a
very good test of the SM. Furthermore the measurement
of the cross section is the first step in understanding
any selected $t \bar t$ sample. If physics beyond the SM exists, 
it can change the overall production rate or the rate in different channels.
Finally top quark pair production is also a background 
for searches for new physics.

\subsection{Cross section measurements}
\label{subsec:topairxsection}
The most precise measurement of the $t \bar t$ production 
cross section at the Tevatron is obtained in the lepton+jets 
channels. Both CDF and D0 have performed measurements based 
on kinematical approaches and b-quark jet tagging.
The CDF kinematical approach uses pre-tagged samples 
and a neural network discriminant. CDF measures the ratio 
$\sigma_{t \bar t}/\sigma_{Z/{\gamma^{*}} \rightarrow l^{+}l^{-}}$ of 
the $t \bar t$ cross section and the Z~boson production cross section
which allows to reduce the systematic uncertainties from luminosity. 
By using the theoretical value for  $\sigma_{Z/{\gamma^{*}}}$, one can 
extract the $t \bar t$ cross section. Combining this approach with
b-quark jet tagging, CDF finds 
$\sigma_{t \bar t} = 7.70 \pm 0.52$~(stat+syst)~pb using a dataset
of 4.3~fb$^{-1}$ and assuming $m_{t}=172.5$~GeV~\cite{cdfttXsection}.
From a combination of a kinematical approach and b-quark jet tagging
D0 determines the $t \bar t$ production cross~section and the background 
from W~boson + heavy or light flavor production simultaneously.
D0 finds 
$\sigma_{t \bar t} = 7.78 \pm 0.25$~(stat)~$^{+0.73}_{-0.59}$~(syst)~pb 
using a dataset of 5.3~fb$^{-1}$ and 
assuming $m_{t}=172.5$~GeV~\cite{d0ttXsection}.
The dominant systematic uncertainties are coming from
jet energy scale, b-quark jet tagging acceptances,
and the estimation of the W+b background.
All measurement are now dominated by systematic
uncertainties.
Combining all their measurement channels, CDF finds 
$\sigma_{t \bar t} = 7.50 \pm 0.48$~(stat+syst)~pb
i.e. a precision of 6.4 \%, with datasets
up to 5.7~fb$^{-1}$. The results are in agreement with the predictions
of the SM and are consistent
across channels, methods and experiments. 

\subsection{Forward backward asymmetry}
\label{subsec:afb}
At the Tevatron the $t \bar t$ production is predicted to be charge 
symmetric at LO in QCD. However NLO
calculations predicts asymmetries in the 5\%-10\% range~\cite{asym} 
and next-to-next-to-leading order (NNLO) 
calculations predict significant corrections for $t \bar t$ production 
in association with a jet~\cite{dittmaier}.
The charge asymmetry arises from interferences between symmetric and 
antisymetric contributions
under the exchange $t \leftrightarrow \bar t$. 
The charge asymmetry depends 
on the region of phase space
and, in particular, on the production of an additional jet. 
Tree level and box diagrams intereferences give a positive asymmetry
while initial and final state radiation intereferences give a negative 
asymmetry.
The small asymmetries expected in the SM makes this
a sensitive probe for new physics~\cite{afbnp}.

Experimentally the $t$ direction and rapidity,
defined as function of the polar angle $\theta$ and the ratio of 
the particle's momentum to its energy $\beta$ as
$ y(\theta ,\beta ) = {1 \over 2} ln [(1+\beta cos \theta ) / 
(1-\beta cos \theta) ] $,
are reconstructed and the rapidity difference $\Delta y = y_t - y_{\bar t}$
is used. The background
is then subtracted from data. CDF estimates the background 
from Monte Carlo (MC) simulations. For D0 the background is fitted with 
likelihood discriminants.

The raw asymmetry 
\begin{eqnarray}
A_{fb} = {{N(\Delta y >0)-N(\Delta y<0)} \over  {N(\Delta y>0)+N(\Delta y<0)}},
\label{eq:d0-afb1}
\end{eqnarray}

where $N(\Delta y >0)$ ($N(\Delta y <0)$) is the number of event with 
positive (negative) $\Delta y$,
is then extracted. The next step consists in performing unfolding
i.e. correcting
the raw asymmetry for acceptance and resolution back to
production level.

Using a dataset of about 5~fb$^{-1}$ both CDF and D0 find that inclusive
asymmetries exceed the SM prediction by 1.5 to 2 standard
deviation~\cite{cdfasym1,d0asym1} with an unclear dependence on the 
invariant mass of the $t \bar t$
system i.e. $M_{t \bar t}$, and  $\Delta y $. Remarkably enough CDF finds
an asymmetry which is about 3 standard deviations away
from the prediction of MC@NLO~\cite{mcatnlo} 
for $M_{t \bar t} > 450$~GeV~\cite{cdfasym1}.  

CDF updated recently its forward backward asymmetry measurement with the full
datasets 8.7~fb$^{-1}$ in the lepton+jets channels
and using POWHEG~\cite{powheg} (with electroweak corrections) for the SM 
predictions.

%%%%%%%%%%%%%%%%%%%%%%%%%%%%%%%%%%%%%%%%%%%%%%%%%%%%%%%%%%%%%%%%%%%%%%%%%
%%
\begin{figure}[hbt]
\begin{center}
\epsfig{file=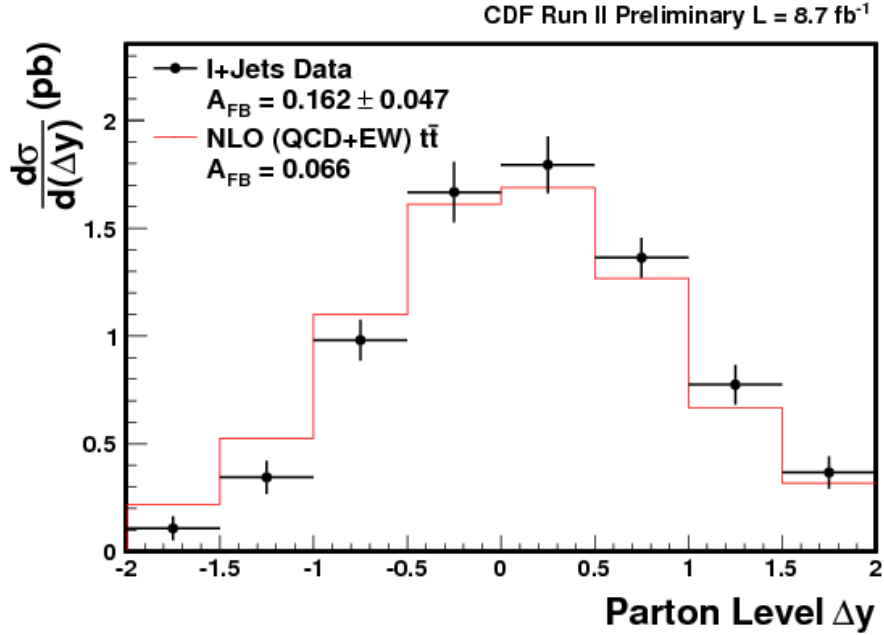,height=3.5in}
\caption{$\Delta y$ distribution from CDF
after background subtraction and unfolding (from~\cite{cdfasym2}).}
\label{fig:asymcdf1}
\end{center}
\end{figure}
%%%%%%%%%%%%%%%%%%%%%%%%%%%%%%%%%%%%%%%%%%%%%%%%%%%%%%%%%%%%%%%%%%%%%%%%%%%

\begin{figure}[hbt]
\begin{center}
\epsfig{file=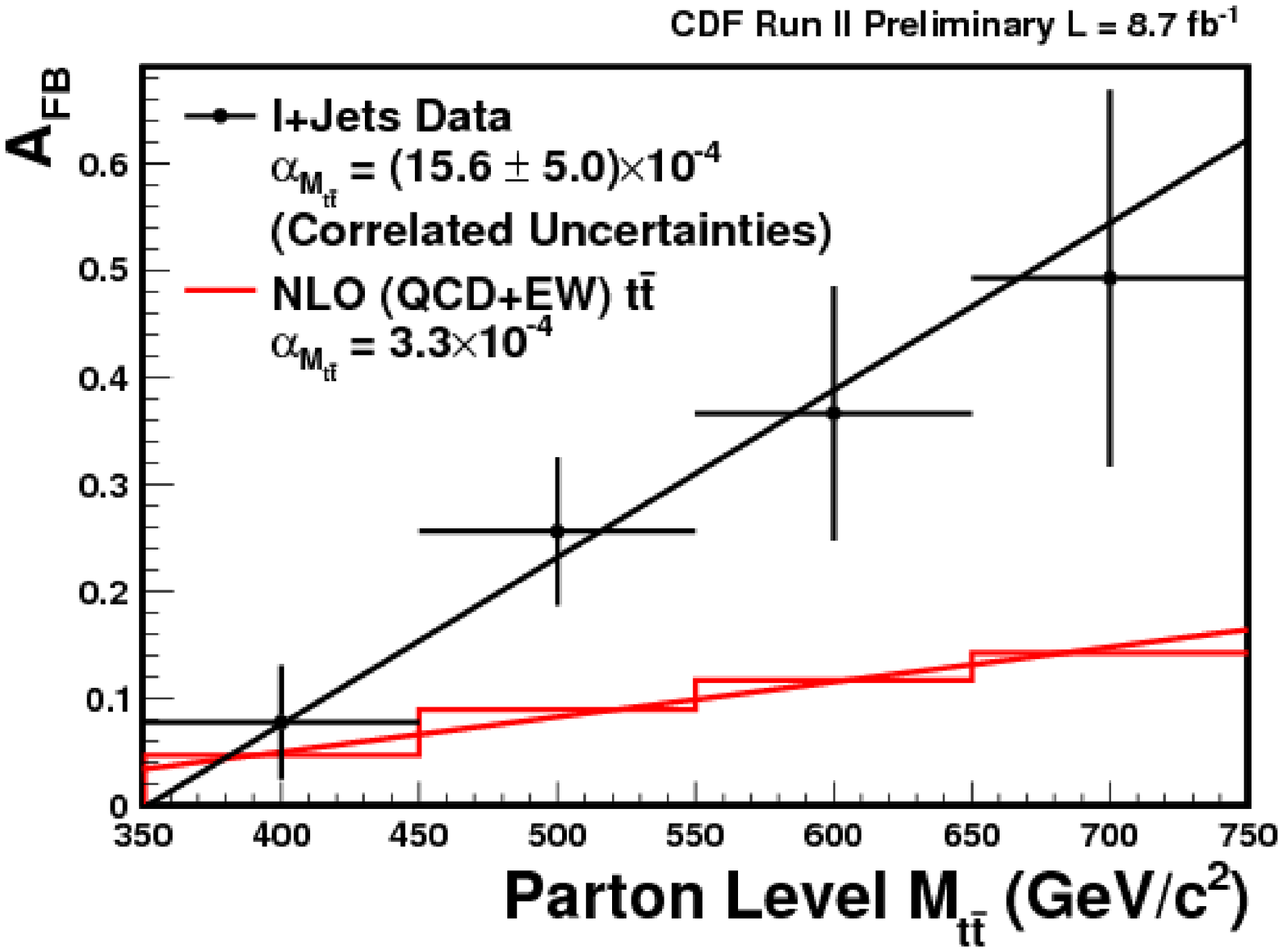,height=3.5in}
\caption{$A_{FB}$ as a function of $M_{t \bar t}$ from CDF data with
a dataset of 8.7~fb$^{-1}$ (from~\cite{cdfasym2}).}
\label{fig:asymcdf2}
\end{center}
\end{figure}

Fig.~\ref{fig:asymcdf1} shows the $\Delta y$ distribution from CDF
after background subtraction and unfolding. CDF finds an inclusive
asymmetry~\cite{cdfasym2}:

\begin{eqnarray}
A_{fb} = 0.162 \pm 0.041 (stat)~\pm 0.022 (syst)
\label{eq:cdfafb1}
\end{eqnarray}

which is compatible with their previous measurement
with a smaller dataset. In the SM the forward backward 
asymmetry is expected to increase with $M_{t \bar t}$ 
and with $\Delta y$~\cite{asymdep}. Beyond SM physics could 
show different dependences than in the SM.
As can be seen from Fig.~\ref{fig:asymcdf2}
the $M_{t \bar t}$ dependency is
well fit with a linear function. The same holds for the $\Delta y$
dependence~\cite{cdfasym2}.

%As can be seen from Fig.~\ref{fig:asymcdf2} and~\ref{fig:asymcdf3}
%the $M_{t \bar t}$ and $\Delta y$ dependencies are
%well fit with a linear functions. The slope parameters are measured 
%and compared with prediction as shown in Tab.~\ref{tab:cdf1}

%%%%%%%%%%%%%%%%%%%%%%%%%%%%%%%%%%%%%%%%%%%%%%%%%%%%%%%%%%%%%%%%%%%%%%%%%
%%
%%%%%%%%%%%%%%%%%%%%%%%%%%%%%%%%%%%%%%%%%%%%%%%%%%%%%%%%%%%%%%%%%%%%%%%%%%%

\begin{table}[htb]
\begin{center}
\begin{tabular} {|c|c|c|}  \hline
slope parameter & $A_{FB }$ vs $M_{t \bar t} $ &  
 $A_{FB }$ vs $\Delta y$ \\ \hline
DATA & $(15.6 \pm 5.0) \times 10^{-4} $ & 
$(30.6 \pm 8.6) \times 10^{-2} $ \\ \hline
SM & $ 3.3  \times 10^{-4} $ & $ 10.3 \times 10^{-2} $ \\ \hline
\end{tabular}
\caption{CDF $t \bar t $ forward backward asymmetry as a function of
$M_{t \bar t}$ and $\Delta y$ : best fit slope for observed data compared 
to NLO prediction (from~\cite {cdfasym2}).}
\label{tab:cdf1}
\end{center}
\end{table}

Tab.\ref{tab:cdf1} shows the values of the slope parameter of the
best linear fit for the dependences of $A_{FB}$ on $M_{t \bar t}$ and
${\Delta y}$ for both data and SM predictions.

\subsection{Spin correlations}
\label{subsec:spincorr}
Top quark pairs are produced with a definite
spin state depending on the production mechanism
i.e. spin 1 for $q \bar q $ annihilation and
spin 0 for gluon fusion. Since $t$ decays before 
hadronizing the spin information is expected to 
pass to the decay products.
The spin correlation from $t$ decay 
products can be measured from the angular distribution
\begin{eqnarray}
{1 \over \sigma} {d^2 \sigma \over  {d \cos \theta_1 \theta_2}} =
{1 \over 4 }(1-C cos \theta_1 cos \theta_2 )
\label{eq:d0spin1}
\end{eqnarray}
where $\theta_1$ and $\theta_2$ denote the angles between
the direction of flight of the decay leptons (for leptonically
decaying W bosons) or jets (for hadronically decaying W bosons)
in the parent $t$ and $\bar t$ rest frames and the spin quantization
axis, and introducing the correlation strength $C$ :
\begin{eqnarray}
C = {{ N_{\uparrow \uparrow} +  N_{\downarrow \downarrow}
- N_{\uparrow \downarrow} -  N_{\downarrow \uparrow} }
\over
{ N_{\uparrow \uparrow} +  N_{\downarrow \downarrow}
+ N_{\uparrow \downarrow} +  N_{\downarrow \uparrow} }}
\label{eq:d0spin2}
\end{eqnarray}
where $N_{\uparrow \uparrow}$ ($N_{\uparrow \downarrow}$ etc...)
is the number of $t \bar t$ pairs with the spins parallel (or 
anti-parallel etc...) to a certain basis.
In the beam momentum vector as the spin quantization axis, the NLO 
SM prediction
for the correlation strength is $C = 0.777^{+0.027}_{-0.042}$~\cite{spincorrth}.

At the Tevatron the spin correlations are measured
using templates with different C values. The templates are then compared
with data using maximum likelihood fits.
Using datasets of 5.1 and 5.4~fb$^{-1}$ respectively, 
CDF finds
$C=0.04 \pm 0.56$~(stat+syst)~\cite{cdfspincorr1} 
and D0 finds
$C=0.10 \pm 0.45$~(stat+syst)~\cite{d0spincorr1} 
both in the dilepton channel. In the lepton+jets and using a dataset
of 5.3~fb$^{-1}$ CDF finds
$C=0.72 \pm 0.69$~(stat+syst)~\cite{cdfspincorr2}.
The results are consistent with SM
expectations but are limited by statistical 
uncertainties.

The D0 experiment has developped an alternate method 
based on the so called matrix element method where
an event probability can be defined from the differential
cross section of the $t \bar t$ production process $\sigma(y,H)$, 
the detector response
$W(x,y)$ that describes the probability of a partonic final state $y$
to be measured as $x$ and the parton density functions together with the
correlation (H=1) and no correlation (H=0) hypotheses~\cite{d0spincorr2}:
\begin{eqnarray}
P(x,H) \sim  \int d^6\sigma(y,H) W(x,y) f_{PDF}(q_1) f_{PDF}(q_2) dq_1 dq_2 . 
\label{eq:d0spin3}
\end{eqnarray}

One can built a discriminating variable R from these
probabilities:
\begin{eqnarray}
R = {P(x, H=1) \over {P(x,H=1) + P(x,H=0) }} .
\label{eq:d0spin4}
\end{eqnarray}

\begin{figure}[hbt]
\begin{center}
\epsfig{file=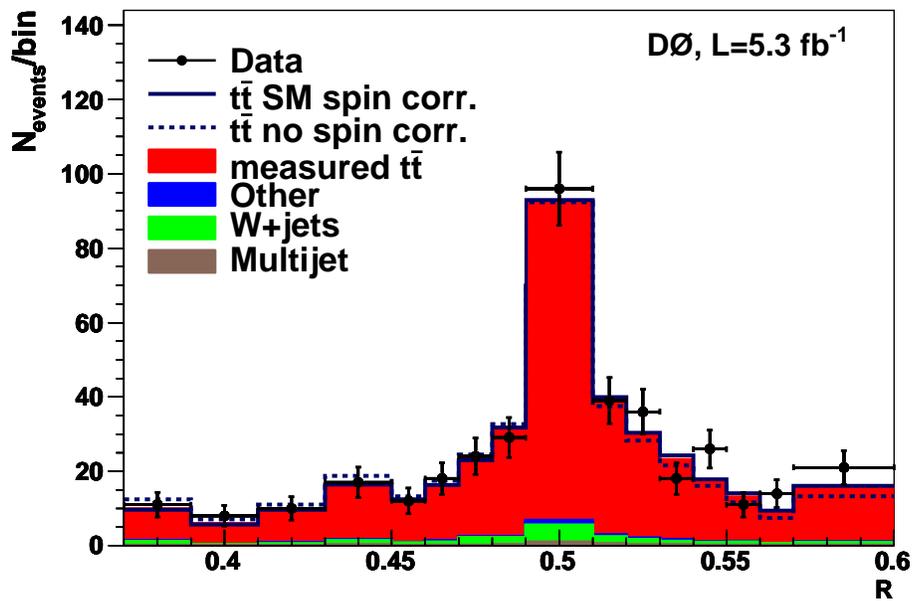,height=3.5in}
\caption{D0 discriminating variable R (see text) 
for the $t \bar t$ spin correlations
measurements with the matrix element method (from~\cite{d0spincorr2}).}
\label{fig:d0spincorr2}
\end{center}
\end{figure}

Fig.~\ref{fig:d0spincorr2} shows the distribution of the discriminant 
variable R from D0 in the lepton+jets channel using a dataset of 
5.3~fb$^{-1}$~\cite{d0spincorr2}.
A binned maximum likelihood fit to the R distribution is performed
to extract the correlation strength.
D0 finds $C^{dil}=0.57 \pm 0.31$~(stat+syst)
in the dileptons channels
and $C^{l+jets}=0.89\pm0.33$~(stat+syst) 
in the lepton+jets channels
using datasets of 5.3 and 5.4~fb$^{-1}$ respectively.

%%%%%%%%%%%%%%%%%%%%%%%%%%%%%%%%%%%%%%%%%%%%%%%%%%%%%%%%%%%%%%%%%%%%%%%%%
%%
%%%%%%%%%%%%%%%%%%%%%%%%%%%%%%%%%%%%%%%%%%%%%%%%%%%%%%%%%%%%%%%%%%%%%%%%%%%

Combining the lepton+jets and the dilepton channels allows the D0
experiment to find $C=0.66 \pm 0.23$~(stat+syst) and $C > 0.26$ at
95 \% of confidence limit as well as exclude the C=0 hypothesis at the level
of 3.1 standard deviations. This is the first evidence of non zero
spin correlation in $t \bar t$ production~\cite{d0spincorr2}.

\subsection{Ratio of branching ratios}
\label{subsec:rb}

In the SM the ratio of branching fractions defined by:

\begin{eqnarray}
R_b = { {BR(t \rightarrow Wb)} \over {BR(t \rightarrow Wq)}}
= {{|V_{tb}|^2}    \over  
{| V_{td}|^2  + |V_{ts}|^2  + |V_{tb}|^2}}
\label{eq:rb1}
\end{eqnarray}
is constrained by the unitarity of the CKM matrix to be equal to 1.
Any value below 1 could indicated new physics.
One can drop the assumption $R=1$ in the $t \bar t$ cross section
measurements. This has been done 
by the D0 experiment for the 
dilepton and lepton+jets channels using a dataset of 
5.4~fb$^{-1}$~\cite{d0rb1} and by the the CDF experiment in the
lepton+jets channels using a dataset
of  8.7~fb$^{-1}$~\cite{cdfrb1}.
The measurement of CDF is based on the determination of the number 
of b-quark jets in $t \bar t$ events using lepton+jets samples i.e.
samples with at least 3 jets in the final state further divided into 
subsets according to the lepton type, jet multiplicity and number of 
b-tagged jets. 
The comparison between the prediction and the observed data in each
subsample is made by using a likelihood function where $R_b$ and the
$t \bar t$ production cross section are simultaneously fit.
CDF measures $R=0.94 \pm 0.09$~(stat+syst) and 
$\sigma_{t \bar t} = 7.5 \pm 1.0$~(stat+syst)~pb. Assuming the unitarity
of the CKM matrix and 3 generations of quarks one can extract 
$V_{tb} = 0.97 \pm 0.05$~(stat+syst).
Using a dataset of 5.4~fb$^{-1}$ and combining the lepton+jets
and the dilepton channels, the D0 experiment measures~\cite{d0rb1}
$R=0.90 \pm 0.04$~(stat+syst), 
$\sigma_{t \bar t} = 7.74^{+0.67}_{-0.57}$~(stat+syst)~pb. 
Assuming the unitarity of the CKM matrix and 3 generations of quarks, 
D0 extracts 
$V_{tb} = 0.95 \pm 0.02$~(stat+syst) and $V_{tb} > 0.88$ at 95\% confidence
level. As we will see in the next section, $V_{tb}$ can be directly measured
from the measurement of the single $t$ production cross~section.

\section{single top quark production}
\label{sec:singletop}
Single $t$ production via electroweak interactions has been 
observed for the first time by the CDF and DO experiments 
in march 2009~\cite{singletop1st}. 
As mentioned above,
single $t$ production is expected to occur via 
electroweak interactions either in the s-channel (33 \%) or the t-channel 
(67 \%)
with the following SM predictions for the cross sections   
$\sigma^{s-channel} = 1.05 \pm 0.07$~pb and
$\sigma^{t-channel} = 2.10 \pm 0.19$~pb
for $m_{t}=172.5$~GeV~\cite{singletop}.
The single $t$ associated production $Wt$ having a production 
cross~section of the order of 0.2~pb is too small at the Tevatron.
Single $t$ production measurement allows for a
direct access to the W-t-b vertex and thus to a
direct measurement of the $V_{tb}$ matrix element of the CKM matrix.
One of the major difficulty with single $t$ production
measurements comes from the fact that there is
a large background with uncertainties larger
than the single $t$ signal itself. As a consequence, the use
of multivariate techniques (MVA) are mandatory and several of them
are used such as boosted decision trees~\cite{bdtref} (BDT), bayesian 
neural network~\cite{bnnref} (BNN), neuroevolution of augmented 
topologies~\cite{neatref} (NEAT).

Using a dataset corresponding to an integrated
luminosity of up to 3.2~fb$^{-1}$ the combination
of Tevatron results gives~\cite{cdfd0singlecombo} 
$\sigma_{single-top} = 
2.76^{+0.58}_{-0.47}$~(stat+syst) for $m_t = 170$~GeV, 
$V_{tb}=0.88 \pm 0.07$~(stat+syst)
and  $V_{tb} > 0.77$ at 95\% confidence level.
These measurements have been updated with larger datasets and, in 
the following, we are going to summarize these new results.

\subsection{Cross section and $V_{tb}$ measurements}
\label{subsec:singletxsec}

The D0 experiment has updated its results in the lepton+jets 
channel~\cite{d0single2}
where the W~boson from the parent $t$ is decaying leptonically
(electron and muon are the lepton flavors under consideration) and 
this with a dataset of 5.4~fb$^{-1}$.
The discriminating variables (single object kinematics, global event 
kinematics, jet reconstruction, $t$ reconstruction and angular correlations) 
allowing to separate the signal from
the background are combined into various MVAs i.e. BDT, BNN and NEAT.
The MVAs are trained separately for the two single $t$ production 
channels namely the s-channel (tb) production considered as the signal 
and the t-channel (tbq) production as part of the background on the one
hand and vice and versa on the other hand. The correlation among the 
outputs of the individual MVA is about 70\%. A second BNN is used to 
construct a combined discriminant for each channel.
A bayesian approach is then used to extract the production cross~section.
The method consists of forming binned likelihood as a product of the six 
analysis channels built from the jet  multiplicities i.e. 2, 3 or 4 jets 
with 1 or 2 b-tagged jets and bins using the full discriminant outputs.
Poisson distribution for the number of events in each bin as well as
uniform prior probabilities for non-negative values of the signal 
cross~sections are assumed. Systematic uncertainties and their correlations
are taken into account assuming a Gaussian prior for each source of
systematic uncertainty. The production cross section is then 
given by the position of the maximum of a posterior probability density 
function of single $t$ cross~section. 
The D0 experiment finds 
$\sigma_{s+t} = 3.43^{+0.73}_{-0.74}$~pb,
$\sigma_{t-channel} = 2.86^{+0.69}_{-0.63}$~pb
and
$\sigma_{s-channel} = 3.43^{+0.38}_{-0.35}$~pb with combined
statistical and systematic uncertainties.

This measurement allows
for a direct measurement of the $V_{tb}$ matrix element 
of the CKM matrix as the single $t$ production cross~section
is directly proportional to $|V_{tb}|^2$.
One can measure $V_{tb}$ assuming V-A couplings
but without assuming 3 generations or the
unitarity of the CKM matrix. One can also maintain
the possibility for an anomalous strength of the
left handed W-t-b coupling $f^L_1$ which would rescale
the single $t$ cross~section~\cite{kane}.
From a bayesian analysis the D0 experiment finds
$|V_{tb} f^L_1 | = 1.02^{+0.10}_{-0.11}$.
Assuming $f^L_1 = 1$ and restricting the prior
to the [0,1] interval one finds $V_{tb} > 0.79$
at the 95\% confidence level.

%%%%%%%%%%%%%%%%%%%%%%%%%%%%%%%%%%%%%%%%%%%%%%%%%%%%%%%%%%%%%%%%%%%%%%%%%
%%
\begin{figure}[hbt]
\begin{center}
\epsfig{file=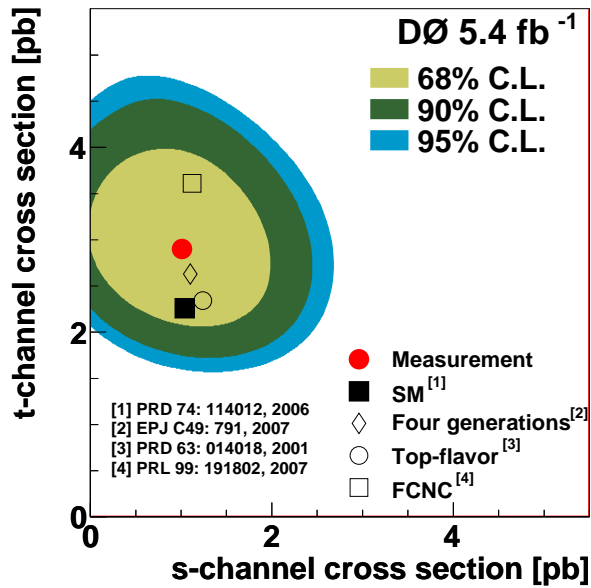,height=3.5in}
\caption{Posterior probability density for the t-channel vs s-channel 
single $t$ production in contours of equal probability density 
(from~\cite{singled02.5}).}
\label{fig:d0single}
\end{center}
\end{figure}
%%%%%%%%%%%%%%%%%%%%%%%%%%%%%%%%%%%%%%%%%%%%%%%%%%%%%%%%%%%%%%%%%%%%%%%%%%%

The D0 experiment has also performed a model independent
measurement of t-channel and s-channel single $t$
cross~sections production. Following the approach 
of~\cite{singled02.5} and using the method described above,
one can construct a two-dimensional posterior probability 
density function as a function of the cross~sections of the 
t- and s-channel processes. Fig.~\ref{fig:d0single} shows
the posterior probability density for the t-channel vs s-channel 
single $t$ production in contours of equal probability density.
The t-channel cross~section is
then extracted from a one-dimensional posterior probability
density function by integrating the previous two-dimensional 
probability density function over the s-channel cross~section 
axis thus making no assumptions about the value of the s-channel 
cross section. One can similarly extract the s-channel cross~section
by integrating the two-dimensional probability density function
over the t-channel cross~section axis.
With a dataset of 5.4~fb$^{-1}$ D0 finds
$\sigma_{t-channel} = 2.90 \pm 0.59$~pb
and 
$\sigma_{s-channel} = 0.98 \pm 0.63$~pb,
statistical and systematic uncertainties being combined.
This gives the most precise cross~sections measurement
in the t-channel with a significance bigger than
5 standard deviations.

\begin{figure}[hbt]
\begin{center}
\epsfig{file=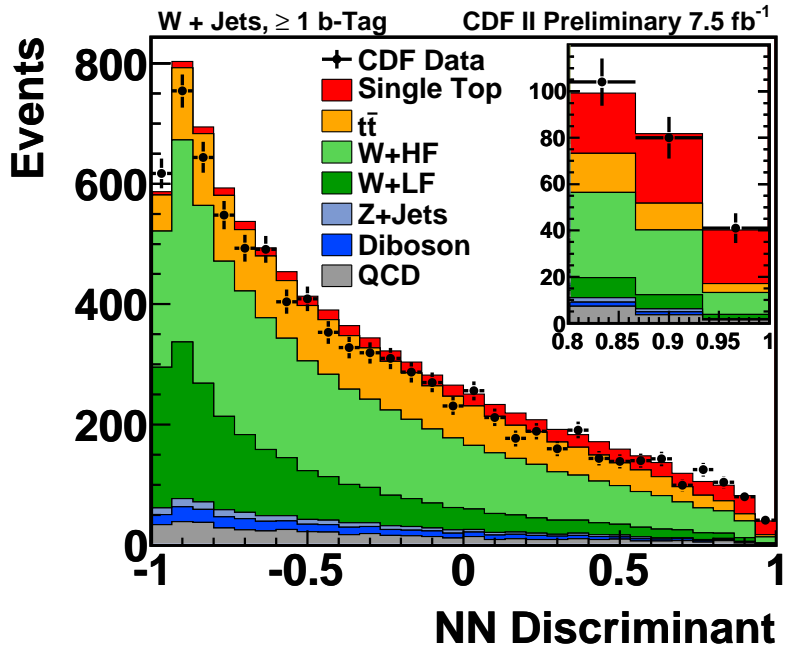,height=3.5in}
\caption{CDF neural network output for the single $t$ production 
cross~section measurement
with the 7.5~fb$^{-1}$ dataset (from~\cite{cdfsingle2}).}
\label{fig:cdfsingle7.5}
\end{center}
\end{figure}

The CDF experiment has also updated its single-top
measurements in the lepton+jets channel
using a dataset of 7.5~fb$^{-1}$\cite{cdfsingle2}.
CDF uses a neural network discriminant with the same 
input variables as the observation analysis. CDF
uses now the NLO POWHEG program for the simulation
of the single $t$ signal.
Fig.~\ref{fig:cdfsingle7.5} shows the neural network
discriminant distribution.

%%%%%%%%%%%%%%%%%%%%%%%%%%%%%%%%%%%%%%%%%%%%%%%%%%%%%%%%%%%%%%%%%%%%%%%%%
%%
%%%%%%%%%%%%%%%%%%%%%%%%%%%%%%%%%%%%%%%%%%%%%%%%%%%%%%%%%%%%%%%%%%%%%%%%%%%

CDF finds 
$\sigma_{single-top} = 3.04^{+0.57}_{-0.53}$~pb for
$m_t = 172.5$~GeV.
Using:
\begin{eqnarray}
|V_{tb}|^2_{measured} =
{\sigma^{measured}_{s+t} \over \sigma^{SM}_{s+t}} |V_{tb}|^2_{SM}
\label{eq:cdfvtb}
\end{eqnarray}
where $|V_{tb}|^2_{SM} \approx 1 $.
CDF extracts 
$V_{tb} = 0.96 \pm 0.09$~(stat+syst) $\pm 0.05$~(theory)
and sets $V_{tb} > 0.78$ at 95\% confidence level.

\section{Search for new physics with top quark }
\label{sec:bsm}

Many searches for physics beyond the SM can be performed with
$t \bar t$ final states at the Tevatron. In particular searches
for new resonances decaying into $t \bar t$ allows to explore
many theories beyond the SM.

Both the CDF and D0 experiments looked for a resonant $t \bar t $ 
production in the lepton+jets channels.

CDF examined the $t \bar t$ invariant mass spectrum of candidate
events where the event kinematics have been reconstructed applying
the matrix element method for SM $t \bar t$ production and decay.
The observed $t \bar t$ invariant mass spectrum is then compared
to templates models of signal such as heavy vector boson decaying
into $t \bar t$ ($Z' \rightarrow t \bar t$) and background processes
in an unbinned maximum likelihood fit. One can then constrain the
$Z' \rightarrow t \bar t$ cross~section times branching ratio.
Using a dataset of 4.8~fb$^{-1}$ CDF data indicate no evidence
of resonant production of $t \bar t$. CDF excludes a benchmark 
model of leptophobic $Z' \rightarrow t \bar t$ with
$m_{Z'} < 900$~GeV at 95\% confidence level~\cite{cdfsrch1}.
%assuming that the $Z'$ width 
%is 1.2 \% of the pole mass.

In the case of the D0 experiment, the $ t \bar t $ invariant mass 
spectrum is reconstructed from the event kinematics. In this
approach the momentum of the neutrino is determined by equating 
the neutrino transverse momentum to the measured missing transverse 
momentum of the event constraining the invariant mass of the charged 
lepton-neutrino system to the W~boson mass and choosing the 
smaller solution of the resulting quadratic equation for the
neutrino momentum longitudinal component along the beam direction.
The reconstructed invariant  $t \bar t$ mass is used to test for
the presence of a signal in the data and to set constraints on the
production cross~section of a narrow $t \bar t $ resonance times 
branching fraction to $t \bar t$ as a function of its mass.
Using a dataset of 5.3~fb$^{-1}$ D0 data indicate no evidence
of resonant production of $t \bar t$. D0 excludes a benchmark 
model of leptophobic $Z' \rightarrow t \bar t$ with
$m_{Z'} < 835$~GeV at 95\% confidence level~\cite{d0srch1}.
    
As mentioned in section~\ref{subsec:afb}, CDF results on the $t \bar t$
forward backward asymmetry $A_{FB}$ indicate a discrepancy with current 
SM predictions. A wide class of models have been built to explain such 
a discrepancy involving the production of a new heavy particle enhancing
 $A_{FB}$. This new heavy particle can also be singly produced in association
with a top quark (or anti-top quark) and further decay into an anti-top quark
and an additional quark (or top quark and an additional quark) looking like
a $t$+jet resonance in $t \bar t $~+~jet events. 
Using a dataset of 8.7~fb$^{-1}$, CDF has performed a search for such events
using the lepton+jets channel with at least 5 jets and at least one b-tagged
jet. CDF finds the data to be consistent with the SM prediction and sets
cross~section upper limits from 0.61~pb to 0.02~pb for resonances ranging
from 200~GeV to 800~GeV~\cite{cdfsrch2}.

\section{Summary}
\label{sec:summary}
The CDF and D0 experiments at the Tevatron provide precision measurements for
$t \bar t$ production cross~section. In most cases the measurements are
now limited by systematic uncertainties.
Forward backward asymmetry of top events keeps indicating a discrepancy
with current NLO QCD prediction.
The Tevatron provides the first evidence for non zero spin correlations
in $t \bar t$ events.
New results on the measurement of single $t$ production cross~sections
reach precisions better than 20\%. The production via the t-channel process
has been observed. Measurements and limits on the matrix element $V_{tb}$
of the CKM matrix have been provided. Finally there is no evidence
for resonant $t \bar t$ production in CDF and D0 data and constraints
have been set on benchmark models such as leptophobic $Z'$.

\end{document}